# MONITORING OF THE PHOTON BEAM

V.I. Alekseev[a], V.A. Baskov[a]*, V.A. Dronov[a], A.I. L'vov[a], A.V. Koltsov[a], Yu.F. Krechetov[b], V.V. Polyansky[a], S. S. Sidorin[a]

[a] *P.N. Lebedev Physical Institute, Moscow, 119991 Russia, 53 Leninsky Ave.*
[b] *Joint Institute for Nuclear Research, 141980 Russia, Moscow Region, Dubna, 6 Joliot-Curie Street*

*E-mail: baskov@x4u.lebedev.ru*

Presented are the characteristics of systems for monitoring the intensity of the bremsstrahlung photon beam of the "Pakhra" accelerator based on Cherenkov counters.

**Introduction**

The calibration channel of quasi-monochromatic electrons was created on the basis of the beam of bremsstrahlung photons for testing and calibration of detectors, electronic equipment on the S-25R electron synchrotron "Pakhra" LPI. It was decided to monitor the photon beam which includes the monitoring of beam intensity and profile at various points of its transport from the inner target of the accelerator to the converter in front of the main spectrometric magnet SP-57 using Cherenkov counters based on solid-state radiators [1]. Monitoring of photon beams in this way was performed earlier in [2-4] works, where monitoring of high-intensity photon beams with energies in the tens of MeV was performed by hodoscope systems using thin optical fibers based on $SO_2$.

The features of using Cherenkov radiation for photon beam monitoring are: 1. low efficiency of photons interaction with material in comparison with the electrons interaction (less than ~$10^3$-$10^4$ times), that makes it possible to use the Cherenkov counter for monitoring photon beams with a wide range of intensities $10^6$ - $10^{10}$ γ/s (noninvasive monitoring system where the photon beam practically does not experience changes [4]); 2. fast time of formation of the Cherenkov pulses in the radiator compared, for example, to the scintillation one (~3-5 times), which is necessary at high beam intensity; 3. radiation resistance of the radiator (quartz glass, organic glass) compared with the scintillator (polystyrene) above. The most important thing is the proportional dependence of the Cherenkov photons number on the number of e⁺e⁻ pairs converted in the substance of the counter, that makes it possible to compare the number of registered e⁺e⁻ pairs to the number of photons passed through it and the total intensity of the photon beam [5,6]. The estimation shows that the number of Cherenkov photons passing through organic glass with a thickness of 1 cm of a photon beam with an intensity of ~$10^9$ γ/s is $N_{chph} \sim 5\times10^7 - 5\times10^8$ photons/s (taking into account the passage of two particles – an electron-positron pair), which is quite reasonable for creating a monitoring system [4,6,7].

**Full intensity monitoring**

The scheme of the calibration quasimonochromatic beam of secondary electrons is shown in Fig. 1. The bremsstrahlung photon beam resulted from electron discharge in the ring to the inner target is formed by lead collimators $K_1 - K_4$ (1) and cleaning magnet (9) after the accelerator chamber output. Then, the beam is transported through the air to the converter (13) located directly on the SP-57 magnet (14). In the magnet there is the separation of electrons coming out of the converter by pulses. The secondary electron beam is formed at an angle φ = 36° relative to the initial photon trajectory using collimators and scintillation counters (16-22).

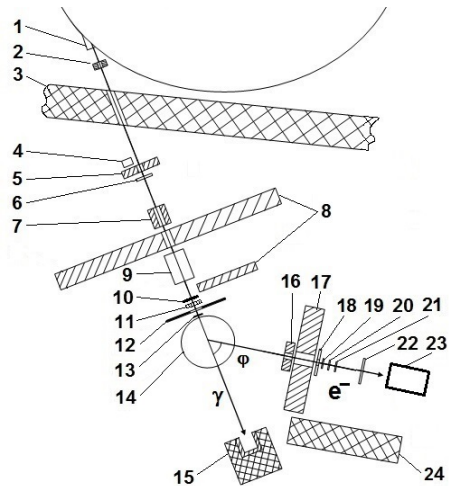

**Fig. 1** The scheme of the calibration quasi-monochromatic beam of secondary electrons: 1 - output window of the accelerator chamber; 2 - collimator $K_1$ (∅ 13 mm); 3 - concrete wall of accelerator hall; 4 - "stretch" monitor; 5 - collimator $K_2$ (hole size 90×90 mm²); 6 - beam monitor; 7 - collimator $K_3$ (∅ 30 mm); 8 - lead protection wall ; 9 - cleaning magnet SP-3; 10 - scintillation counter $S$; 11 – the Cherenkov hodoscope (ChH); 12 - metal plate (hole size 140×140 mm²); 13 - converter; 14 - magnet SP-57; 15 - photon-beam absorber ("burial ground"); 16 - additional collimator (∅ 10 mm); 17 - main collimator $K_4$ (∅ 30 mm); 18 - scintillation anticoincidence counter $A$; 19 - 22 - scintillation counters $S_1$ - $S_4$; 23 - calibrated detector; 24 - concrete block.

The photon beam monitoring system is divided into two subsystems: the first subsystem (6) determines the total intensity of the photon beam at the beginning of the transport channel in front of the $K_3$ collimator (7) with an orifice of ⌀ 30 mm; the second subsystem (10, 11) determines the intensity and position of the photon beam in the horizontal plane in front of the converter (13) located on the poles cut of the SP-57 magnet (14) [1].

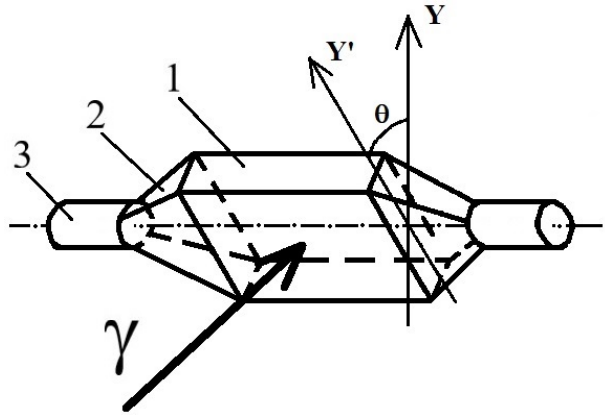

**Fig. 2** Scheme of the bremsstrahlung photon beam monitor located at the beginning of the photon beam channel: 1- organic glass radiator; 2 – air light collector; 3-photoelectric multiplier PMT-85 (Θ - the angle of rotation of the monitor between the vertical Y axis and the working position determined by the y axis Y' (Θ ≈ 40°)).

The monitor that controls the intensity of the photon beam at the beginning of the transport channel (6) (Fig.1) is located in the second experimental hall of the "Pakhra" accelerator directly in the beam behind the $K_2$ collimator and it is a Cherenkov counter (Fig. 2). The counter consists of a radiator (1) based on an organic glass of 100×100×10 mm³ (t = $0.025X_0$ - the counter thickness in radiation lengths $X_0$), from two ends through air light collectors (2) "viewed" by two photoelectronic multipliers PMT-85 (3). The intensity of the photon beam is determined by the number of signals matches from both PMT that registered the conversion $e^+e^-$ pair.

Monitor (6) on the trajectory of the photon beam is turned by an angle Θ ≈ 40° from the relatively to the vertical axis Y (Fig. 2). The angle of total internal reflection for acrylic plastic is about 40° thus, pairing the angle of photon input to the monitor and the angle of total internal reflection of the radiator leads to a decrease in the number of reflections inside the radiator monitor and, accordingly, to the minimum value of the signal attenuation generated by the PMT, which in turn increases the efficiency of the monitor.

Calibration of the monitor was performed directly in the operational position on the photon beam located behind the monitor of the quantometer. When the intensity of the photon beam is ~$2\times10^9$ equivalent photons/s, the count of each monitor arm was ~$5\times10^5$ $e^+e^-$/s, the count of both monitor channels matches was ~$5\times10^3$ matches/s (the voltage on the both PMT dividers was 875 V).

Match signals from the monitor arms were fed to the frequency meter, which was used to visually control the intensity of the photon beam or to the input of the standard "counter" unit of the CAMAC system for recording signals to the computer memory.

The photon beam monitor made it possible to monitor the intensity of each electron beam discharge to the internal target of the accelerator and the total beam intensity for a set time.

### The Cherenkov hodoscope

To monitor the "current" intensity and position of the photon beam on the converter in front of the SP-57 magnet, as well as the total intensity during the set of experimental statistics, a hodoscope based on Cherenkov counters (Cherenkov hodoscope, or ChH) was created. The ChH is also able to determine the formation coordinates of an electron-positron pair and, correspondingly, the point of entry into the magnetic field of SP-57. It will allow determine the angle of the electron (positron) exit from SP-57 and its further trajectory using the established characteristics of the magnetic field and using counters that register secondary electrons (positrons).

The scheme for monitoring of photon beam in front of a copper 1 mm converter with the diameter of 32 mm is shown in Fig. 3. The scintillation counter $S$ of $100\times40\times5$ mm$^3$ is located just a head of the ChH and functions as the converter for each of the ChH counters, and the trigger counter for the system (S + ChH).

The ChH is an assembly consisting of 13 channels, which are transparent and polished on each side by organic glass plates with a thickness of 6.5 mm and a width of 25 mm (insert in Fig. 3). Each plate is rotated by an angle of 90º by heating the vertical axis. At the end of each plate there is the photomultiplier PMT-85. Length of the plate`s part, on which PMT is placed, relative to the point of rotation is defined so that the photocathode was located vertically in the same plane relative to the first photocathode above PMT. The length of the plate parts before and after rotation, on which the first and last PMT is located, is 65 and 70 mm, as well as 570 mm and 165 mm, respectively.

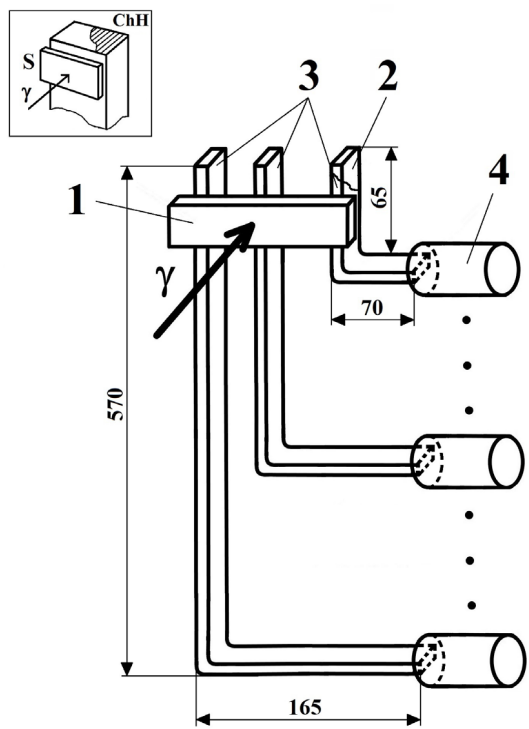

**Fig. 3** Scheme for monitoring the bremsstrahlung photon beam using the Cherenkov hodoscope(ChH): 1 - scintillation counter S; 2 – channel of the Cherenkov hodoscope (ChH) (organic glass plate); 3-metalized Mylar; 4 – PMT–85 (at the insert: *S*-scintillation counter, ChH – Cherenkov hodoscope).

White paper, metallized mylar and metallized foil were studied as a reflecting surface for wrapping the plates of ChH channels. According to research results, it turned out that all three materials give approximately the same results, and metallized mylar was chosen. All sides of the hodoscope`s each plate, except the one to which the non-lubricated PMT is tightly pressed, are wrapped in metallized mylar and black paper.

**Preliminary calibration**

The preliminary calibration ChH was performed by means of ionizing radiation source $^{90}Sr$. The purpose of the calibration was to define the feasibility of using this method in determining the position of the «point» source of charged particles and the preliminary voltage determination on the dividers of the PMT of the ChH.

The calibration scheme is shown in Fig. 4. Electrons from the ionizing radiation source $^{90}Sr$ with a maximum energy of 2.2 MeV passed through the trigger counters $S_1$ and $S_2$ (15×15×1 mm³) and were registered by the ChH channel. The signals from $S_1$, $S_2$ and the ChH channel were fed to the shapers with a constant threshold of $Sh_1$, $Sh_2$ and Sh (the shapers' thresholds were 10 mV). Signals from $Sh_1$ and $Sh_2$ through the delayers $DU_1$ and $DU_2$ were received at the inputs of the coincidence circuit CC. The "Start" signal from the CC was fed to the "start" input of the "counter". At the input "analysis" of the "counter" from the shaper Sh through the delayer DU, the signal was sequentially fed from 1 channel ChH to 13. The collimator K, that is the lead plate of 100×100×5 mm³ with a hole in the center of the plate of ⌀5 mm, was located between the source and the ChH for collimation of the electron flow from the $^{90}Sr$ source.

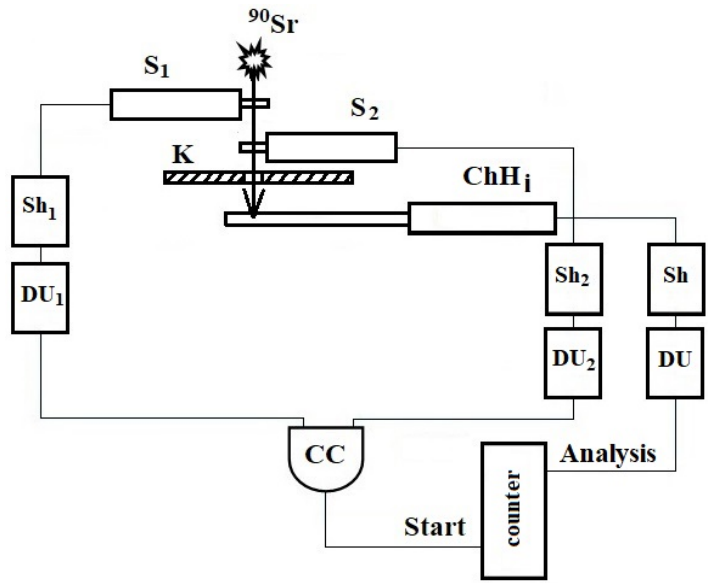

**Fig. 4** Scheme of preliminary calibration of the Cherenkov hodoscope using the $^{90}Sr$ radioactive source: $S_1$ and $S_2$ – scintillation counters; $ChH_i$ – i channel of the Cherenkov hodoscope (ChH); K – collimator; $Sh_1$, $Sh_2$ and Sh – shapers with a constant threshold; $DU_1$, $DU_2$ and DU – delay units; CC – coincidence circuit; "counter" – pulse count counter.

"start" input of the "counter". At the input "analysis" of the "counter" from the shaper Sh through the delayer DU, the signal was sequentially fed from 1 channel ChH to 13. The collimator K, that is the lead plate of $100\times100\times5$ mm$^3$ with a hole in the center of the plate of $\varnothing5$ mm, was located between the source and the ChH for collimation of the electron flow from the $^{90}Sr$ source.

The selection of PMT for use in ChH was not carried out, so the ChH calibration performed in two stages. At the first stage, the signal count of hodoscope`s each channel was sequentially equalized by changing the voltage on the PMT divider when the $^{90}Sr$ ionizing radiation source, lead plate, and counters $S_1$, $S_2$ were located above the corresponding channel. The accounts were equalized relative to the account of the central channel (channel 7), the account of which was taken as the basis. The maximum count of the central channel was determined from the dependence of the triple matches` count of the counter signals $S_1$, $S_2$ and signal 7 of the ChH channel on the voltage divider of the ChH channel 7.

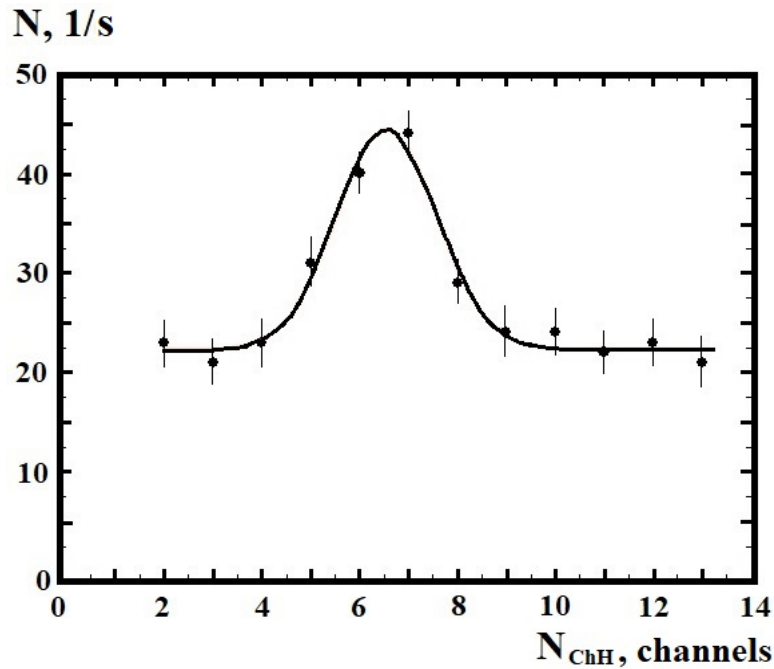

**Fig. 5** The dependence of the events number ($N$) in the channels of the Cherenkov hodoscope ($N_{ChH}$) from the position of the $^{90}Sr$ source.

At the second stage, the profile of the ionizing radiation source was determined by placing the trigger counters $S_1$, $S_2$ and the $^{90}Sr$ source in the absence of a lead collimator in the center of the ChH over the channel 7. The result of the second calibration step is shown in Fig. 5. The figure presents that the maximum account $N_{max} \approx 45$ events/s falls on the central part of the ChH. The full width at the half height of the coordinate beam profile defined by the ChH corresponds to approximately 3 channels of the ChH, which in turn corresponds to 6.5×3=19.5 mm. This is close to the size of the trigger counters, taking into account the multiple scattering of low-energy electrons in the counters and in the air (about 20 mm between the counters $S_1$, $S_2$ and the ChH channel) and the inaccuracy of the trigger counters location relative to each other.

The results of the preliminary calibration using the $^{90}Sr$ source presented the applicability this ChH structure for the detection of charged particles.

**Basic calibration**

The basic calibration of the Cherenkov hodoscope was performed directly on the photon beam in the "calibration" position (Fig. 6a). To obtain the same efficiency of registration electron-positron pairs that results from the conversion of brake photons in the counter $S$ located in front of the hodoscope, the hodoscope was placed on the path of the photon beam in such way that the electron-positron pairs entered the ChH perpendicular to the 25 mm wide side and passed through all channels of the hodoscope. The power supply voltage of the counter S was selected so that the noise count of the PMT counter was no more than ~10 Hz ($U_S$ = 900 V).

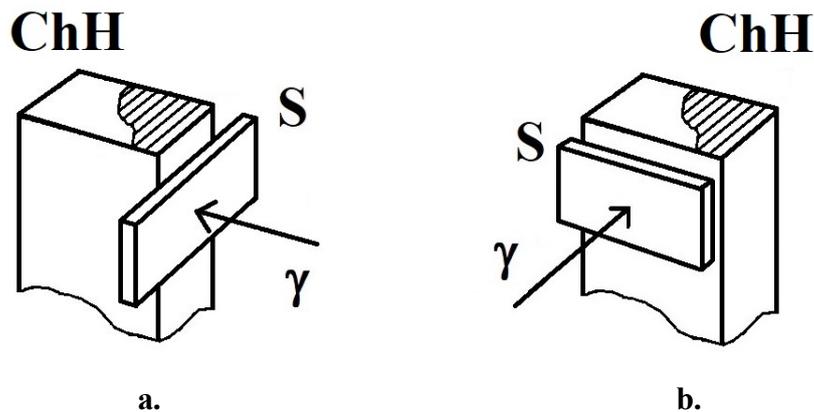

a.   b.

**Fig. 6** The positions of the Cherenkov hodoscope relative to the photon beam: a - position "calibration"; b - position "working" (S – scintillation counter, ChH - Cherenkov hodoscope).

The block scheme of the ChH photon beam registration is shown in Fig. 7 in the positions "calibration" (1) and "working" (2). The signal from the conversion electron-positron pair, which occurred in S from the photon passage through it, was fed to the "splitter" block via the shaper Sh and the delay DU. Then signals from the "splitter" block were fed to the second inputs of the coincidence circuit $CC_1 – CC_{13}$, and the signals formed by the $Sh_1 – Sh_{13}$ shapers from the channels of the hodoscope were fed to the first inputs of $CC_1 – CC_{13}$. Signals from the coincidence circuit $CC_1 – CC_{13}$ were then fed to the inputs of the multi-channel block "registers", which was launched from the "Start" signal that came from the "splitter" block. Using the "splitter" block the "recording" of the signal from the "registers" block was made to the computer memory through the crate controller (CC) of the CAMAC system.

The alignment of the account of each channel hodoscope in the "calibrate" position occurred by varying the voltage on the voltage divider counters of the hodoscope. The dependence of the channel count change on the voltage on the PMT voltage divider was constructed for each ChH channel. Further we selected the equal score value for all channels of the hodoscope, which was 180 ± 10 events/s.

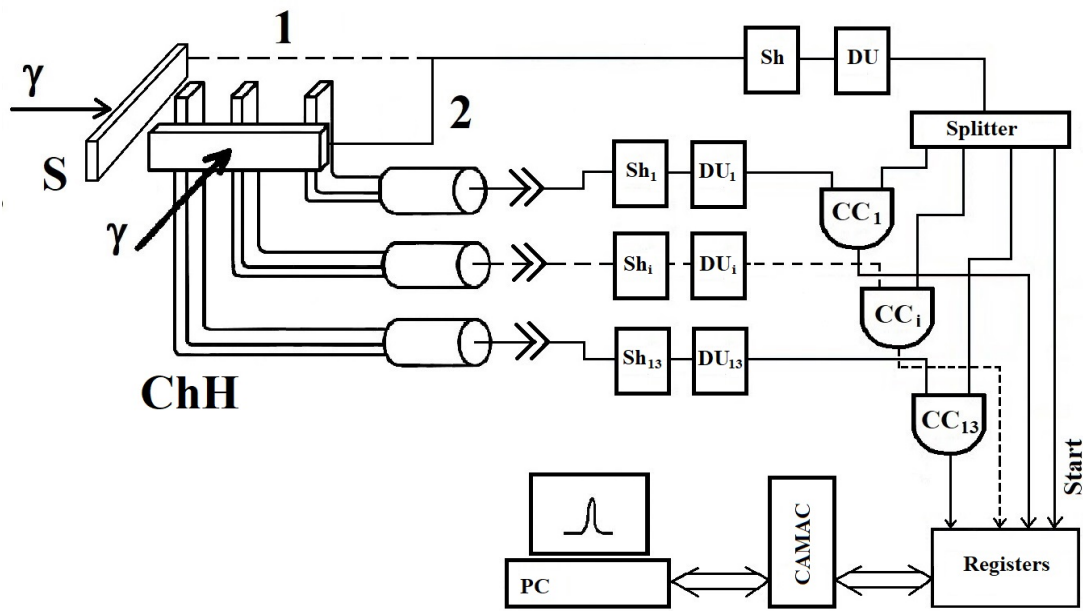

**Fig. 7** Block scheme of the "calibration" position (1) and "working" position (2) of the Cherenkov hodoscope (ChH) on a photon beam: S – scintillation counter; ChH - Cherenkov hodoscope; Sh, $Sh_1 – Sh_{13}$ - shapers; DU, $DU_1 – DU_{13}$ - delays; CC, $CC_1 - CC_{13}$- coincidence circuits; "splitter" – the "splitter" block; "registers" – the "registers" block; CC – crate-controller; PC – personal computer.

After the accounts alignment of all channels, the hodoscope was put in the "working" position (Fig. 6b), where the registration of conversion electron-positron pairs outputted from the counter S was performed by one of the hodoscope channels (Fig. 3). In this position the electron-positron pairs entered the ChH perpendicular to the 6.5 mm wide side. The ChH aperture in this position was 90 mm and 65 mm horizontally and vertically, respectively. However, taking into account the aperture of the counter S, the working aperture of the ChH was 90 and 40 mm horizontally and vertically, respectively.

**Results**

Figure 8 shows the profile of the photon beam in front of the converter located on the cross-section of the poles of the magnet SP-57 (14) (Fig. 1), measured by the Cherenkov hodoscope.

The full width at half the height of the profile was 5 channels, which, taking into account the width of the ChH channel, corresponds to the diameter of the collimator $K_3$ (⌀=30 mm) (7), located in front of the cleaning magnet SP-3 (8) and ChH [1]. The total number of events in the photon beam profile histogram was about ~$2·10^3$ events/s that corresponds to the intensity of the photon beam ~$10^9$ γ/s incident on the copper converter. The profile shows that the photon beam has a background component of about 13% of the main beam total intensity (5 central channels) or ~$1.3·10^8$ γ/sec. Hence it can be seen that the number of events determined by the ChH (~$2·10^3$ events/sec) is significantly less than not only the intensity of the entire photon beam (~$10^9$ γ/sec), but also the intensity of the background component (~$1.3·10^8$ γ/sec).

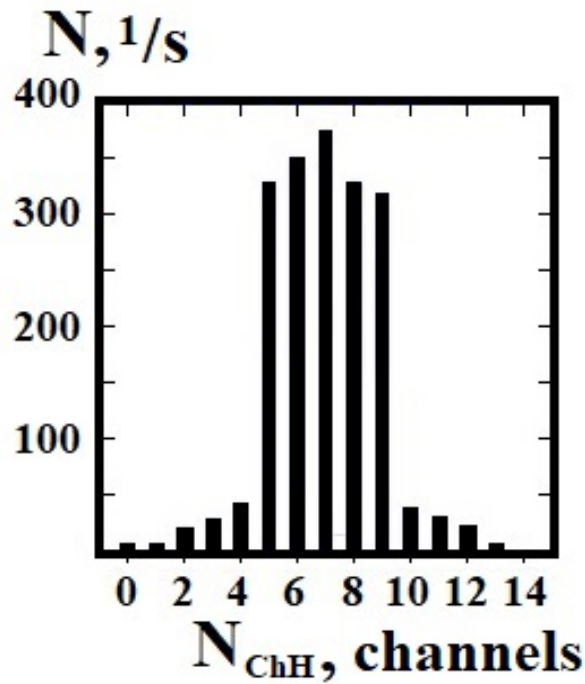

**Fig. 8** The photon beam profile, measured by Cherenkov hodoscope in front of the converter on the cut poles of the magnet SP-57.

**Conclusion**

The presented system for monitoring the braking photon beam, designed to form a calibration quasi-monochromatic beam of secondary electrons, based on Cherenkov radiation is able to control the intensity and profile of the photon beam at various points of its transport.

It should be noted that monitoring systems based on Cherenkov radiation have great design and functionality in experimental practice. For example, within the experimentation by use of photon beam to research the cross section of nuclear reactions, the profile of the photon beam at the input to the experimental target can be determined at each "dump" of the primary beam by the accelerator on the internal tungsten target and, thus, determine the integral profile for a set time. The channels of the Cherenkov hodoscope can be an "active converters", i.e. they can work without a primary converter in front of hodoscope. The signals of the hodoscope counters can be entered into the main trigger that forms a beam of secondary electrons, which will enable to determine the coordinate of the electron (positron) formation and, accordingly, the point of its input into the SP-57 magnetic field. Henceforth the angle of the electron output from the magnet and its further trajectory can be determined by the known characteristics of the magnetic field and geometric parameters of the magnetic system. Knowledge of the electron trajectory will improve (by ~1.5-2 times) the energy resolution of the secondary electron beam [1].

*This work was supported by grants from the Russian Foundation for Basic Research (NICA - RFBR) No. 18-02-40061 and No. 18-02-40079.*